\documentclass[twocolumn]{jpsj2} 
\def\simge{\lower0.7ex\hbox{$\ \overset{>}{\sim}\ $}}
\def\simle{\lower0.7ex\hbox{$\ \overset{<}{\sim}\ $}}
\def\doteqa{\mathrel{\offinterlineskip\setbox0\hbox{$=$}%
\vbox to\ht0{\vss\hsize\wd0%
\hbox to\wd0{\hfil$\smash.$\hfil\hfil\hfil\hfil\hfil}%
\vskip0.1zw\copy0\vskip0.1zw%
\hbox to\wd0{\hfil\hfil\hfil\hfil\hfil$\smash.$\hfil}%
\vss}}}

\newcommand{\dg}{\dagger}

\title{Theory of Ultrasonic Dispersion in Local Phonon Systems Coupled with Conduction Electrons}

\author{Kazumasa \textsc{HATTORI}$^{1,2}$ and Kazumasa \textsc{MIYAKE}$^1$}

\inst{$^1$Division of Materials Physics, Department of Materials Engineering Science, Graduate School of Engineering Science, Osaka University, Toyonaka, Osaka 560-8531, Japan \\
$^2$Institute for Solid State Physics, University of Tokyo, Kashiwa, Chiba 277-8581,Japan
}

\abst{The physical origin of frequency dependence in elastic constants, which are often found in an ultrasound propagation in filled skutterudites and clathrate compounds, is investigated theoretically. This dependence arises from a coupling between the acoustic phonon and some optical phonons, which strongly interact with electrons. Using a self-consistent ladder approximation together with a pseudofermion mapping of the phonon to the single site Holstein Anderson model, a soft mode of the optical phonon at zero frequency is shown to emerge. The temperature dependence of the spectral weight of this soft mode shows an activation-type behavior, which is characterized by the optical phonon frequency. These features can generate the frequency dependence and the shoulder in the elastic constants observed in some filled skutterudites and clathrate compounds.}

\kword{skutterudite, clathrate, ultrasonic dispersion, elastic constant, local phonon, Holstein Anderson model}

\begin{document}
\maketitle
\section{Introduction}
Recently, the frequency dependence of ultrasonic velocity in some filled skutterudites and clathrate compounds has attracted a great deal of attention \cite{GotoNemoto1,GotoNemoto2,GotoNemoto3,NemotoPhysicaB1,NemotoPhysicaB2}. In these systems, it is believed that a rare-earth ion in a rigid cage moves rather independently in the anharmonic potential of off-center minima. In PrOs$_4$Sb$_{12}$, for example, it is expected that these ionic degrees of freedom are important to heavy fermion superconductivity and its large effective quasiparticle mass \cite{Bau}. The data of elastic constants (EC) in these compounds exhibit frequency dependence at temperatures ranging from 10 to 40 K. This temperature range is on the order of the Einstein-like modes of the rare-earth oscillation in these compounds \cite{Keppens, Hermann, tsutsui, Ogita1, Ogita2}. For PrOs$_4$Sb$_{12}$, large softening of acoustic phonons is also reported \cite{iwasa}. Recently, Iwasa {\it et al.}, observed the development of a quasi-elastic peak at low temperatures in Pr-based filled skutterudites \cite{iwasa2}. 

Usually, the frequency dependence of EC manifests as a shoulder in its temperature dependence. This is well fitted by a Debye-type formula:
\begin{eqnarray}
C(z)= C_{\infty}+\frac{C_0-C_{\infty}}{1+(z\tau)^2},\label{eq1}
\end{eqnarray}
where $C(z)$ is the frequency-dependent EC and $C_{\infty}\equiv C(\infty)$ and $C_0\equiv C(0)$ are constants in frequencies. In the present paper, we use $z$ as the frequency of dynamic quantities. The lifetime $\tau$ in eq. (\ref{eq1}) is given as $\tau=\tau_0\exp(E_{\Delta}/T)$, where $E_{\Delta}$ is a characteristic energy scale and $T$ is the temperature. The exponential $T$-dependence gives the shoulder at $z\tau \sim 1$, at approximately $T\sim 40$ K for $E_{\Delta}\sim 200$ K, even though the frequency of the ultrasound is of the order of mK. This is called ``ultrasonic dispersion'' (UD). This form of the relaxation time appears to be related to a kind of thermal activation-type process, so that in ultrasonic experiments it has been interpreted that there is an off-center potential and the origin of the relaxation is due to the thermal hopping between the off-center sites. In this respect, we discussed a possible scenario for the realization of heavy fermions in SmOs$_4$Sb$_{12}$\cite{SmOsSb} in the strong coupling limit of electron-phonon couplings\cite{HattoriHirayamaMiyake1,HattoriHirayamaMiyake2}, which is a natural extension of a two-level system\cite{Kondo1,Kondo2,YuAnderson, MiyakeMatsuura,Vlad}. Since the UD is observed only in a specific mode, it is claimed that the off-center degrees of freedom have a degenerate and anisotropic ground state. On the other hand, the results of a neutron scattering experiment in PrOs$_4$Sb$_{12}$ suggests that there are no anisotropic charge distributions of the Pr-nuclei at low temperatures in particular\cite{Kaneko}.  

In order to discuss the frequency dependence in EC, it is convenient to treat the two frequency regions separately. One typical region is $z\tau\ll 1$, where the time scale of the ultrasound is much longer than that of the relaxation mode. At this stage, the origin of the relaxation mode is unknown. The relaxation mode is scattered in a much shorter time than the oscillation period of the ultrasound. Thus, the strain caused by the ultrasound can be regarded as static. The ultrasound attenuation due to the anharmonicity of the lattice in this region was discussed by Akhieser \cite{Akhieser}, and later by Woodruff and Ehrenreich \cite{Woodruff}, using the semi-classical approach (Boltzmann equation). For the case of $z\tau \gg 1$, the Landau and Rumer theory, which is ``golden rule'' treatment, can be applied \cite{LandauRumer}. Although the former description appears to be good even in $z\tau > 1$ qualitatively, there is no theoretical reasoning to use the former in this region.

Using the Woodruff and Ehrenreich theory, the frequency dependence in the EC was discussed in KTaO$_3$ by Barrett \cite{Barrett} from both theoretical and experimental points of view. In KTaO$_3$, the activation factor $\exp(\omega_0/T)$ also appears in the ultrasound velocity, where $\omega_0$ roughly corresponds to the frequency of the optical mode. This factor arises from the specific heat of the free optical mode through the time-dependent Boltzmann equation. In comparison, the theory for the frequency dependence in filled skutterudite and clathrate compounds is more difficult, since we need to bridge the above two regions continuously and to take into account the electron-phonon couplings. 

In the present paper, we discuss a simple interpretation of this problem. In \S\ref{sec:phenomenolo}, we review the Green function formulation in electron-phonon systems. We then discuss two types of couplings between the optical and acoustic phonons, and give the phenomenological interpretation of the Debye formula used in the ultrasonic experiments. In \S \ref{sec:micro}, we investigate the Holstein-Anderson model at finite temperatures using the self-consistent ladder approximation. We then show the results of the renormalized sound velocity, giving the dispersion in EC, together with the nature of soft optical phonon modes. In \S \ref {Discussion}, we discuss the applicability of the present theory to the real materials and future problems. Finally, in \S \ref{sec:sum}, we summarize the results.

\section{Coupling between acoustic and optical phonons}\label{sec:phenomenolo}
 In this section, we discuss how eq. (\ref{eq1}) is derived in interacting electron-phonon systems. Note that the displacements due to the acoustic modes near ${\bf{q}}=0$ include contributions from optical modes at ${\bf{q}}=0$. This essential point is described in \S \ref{harmoniccoupling}. In \S \ref{anharmoniccoupling}, we discuss the case in which the anharmonicity of the lattice is important.

\subsection{Harmonic coupling}\label{harmoniccoupling}
The classical Hamiltonian for the phonon system in the harmonic approximation is written using a force constant matrix $K$ as
\begin{eqnarray}
H_{\rm ph}\!\!\! &=& \!\!\!\sum_{{\bf i},n, \nu}\frac{P^{\nu}_{{\bf i}n}P^{\nu}_{{\bf i}n}}{2M_n}+\sum_{{\bf i},{\bf j}} \sum_{\mu\nu=x,y,z}\sum_{n,m}\frac{1}{2}K_{{\bf i},{\bf j},n,m}^{\mu\nu}X_{{\bf i}n}^{\mu}X_{{\bf j}m}^{\nu},\nonumber\\
\label{Hpgeneral}
\end{eqnarray}
where $P_{\bf i}^{\mu}$($X_{\bf i}^{\mu}$) is the momentum (displacement) variable of the ion in the $\mu$ direction located at the $n$-th site in the ${\bf i}$-th unit cell. $M_n$ is the mass of the ion at the $n$-th site in a unit cell. Introducing variables $p_{{\bf i}n}^{\mu}\equiv P_{\bf i}^{\mu}/\sqrt{M_{n}}$, $x_{{\bf i}n}^{\mu}\equiv \sqrt{M_{n}}X_{{\bf i}n}^{\mu}$ and $k_{{\bf i},{\bf j},n,m}^{\mu\nu}\equiv \sqrt{M_{n}}^{-1}K_{{\bf i},{\bf j},n,m}^{\mu\nu}\sqrt{M_{m}}^{-1}$, the Hamiltonian (\ref{Hpgeneral}) is reduced to
\begin{eqnarray}
H_{\rm ph}\!\!\! &=&\!\!\! \frac{1}{2}\Bigg(\sum_{{\bf i},n, \nu}p^{\nu}_{{\bf i}n}p^{\nu}_{{\bf i}n}+\sum_{{\bf i},{\bf j}} \sum_{\mu\nu=x,y,z}\sum_{n,m}k_{{\bf i},{\bf j},n,m}^{\mu\nu}x_{{\bf i}n}^{\mu}x_{{\bf j}m}^{\nu}\Bigg) \nonumber\\
\!\!\!&=&\!\!\! \frac{1}{2}\sum_{{\bf q}}\Bigg(\sum_{n, \nu}p^{\nu}_{n}({\bf q})p^{\nu}_{n}(-{\bf q})\nonumber\\
&&+\sum_{\mu\nu=x,y,z}\sum_{n,m}k_{n,m}^{\mu\nu}({\bf q})x_n^{\mu}({\bf q})x_m^{\nu}(-{\bf q})\Bigg),\label{Hphobasis}
\end{eqnarray}
where we have introduced the Fourier components of each variable in the second line. After diagonalizing the $k$-matrix, we obtain all of the eigenvalues in principle as in textbooks\cite{Ziman}. In the usual step of moving to the corresponding quantum Hamiltonian, we use the diagonalized basis. Although we can diagonalize the $k$-matrix and obtain the energy eigenvalues of every phonon branch, we here introduce even and odd parts, $k^e$ and $k^o$, with respect to ${\bf q}$, and concentrate on the low ${\bf q}$ limit of the phonon system. In order to determine the phonon velocity, we need only retain up to the second-order contribution in ${\bf q}$. Thus, we write $k^e=k^{(0)}+k^{(2)}$ and $k^o=k^{(1)}$. Dividing $k$ into $k^e$ and $k^o$ is meaningful only in the region near ${\bf q}=0$. Near ${\bf q}=0$ and $z=0$ ($z$ being the frequency of phonons), $k^o =O(|{\bf q}|)$ acts as the hybridization between some optical and acoustic modes at ${\bf q}\simeq 0$. Since we are interested in the sound velocity of the system under which optical phonons strongly interact with conduction electrons (this comes from the fact that UD has an activation type relaxation), dividing $k$ into the even and odd parts is meaningful. For large $|{\bf q}|\simge a^{-1}$ ($a$ being the lattice constant), this characterization loses its meaning, and $k^o$ should be regarded simply as the hybridization.

In order to discuss the quantum mechanical Hamiltonian, we first diagonalize $k^e$. We define the creation and annihilation operators of the mode $\lambda$ with the energy eigenvalue $\tilde{\omega}_{{\bf q}\lambda}$ (related to the eigenvalue of $k^e$, not $k$) as follows:
\begin{eqnarray}
\tilde{x}_{{\bf q}\lambda}=\sqrt{\frac{1}{2\tilde{\omega}_{{\bf q}\lambda}}} (\tilde{a}_{{\bf q}\lambda}^{\dg}+\tilde{a}_{-{\bf q}\lambda}).
\end{eqnarray}
Here, $\tilde{x}_{{\bf q}\lambda}$ is the displacement operator that diagonalizes $k^{e}$. Using this set of displacement variables, Hamiltonian (\ref{Hphobasis}) can be written as
\begin{eqnarray}
H_{\rm ph} = \sum_{{\bf q}\lambda}\tilde{\omega}_{{\bf q}\lambda}(\tilde{a}_{{\bf q}\lambda}^{\dg}\tilde{a}_{{\bf q}\lambda}+\frac{1}{2})+\frac{1}{2}\sum_{{\bf q}\lambda\lambda'}\tilde{x}_{{\bf q}\lambda}k^o_{\lambda\lambda'}({\bf q})\tilde{x}_{-{\bf q}\lambda'}.
\end{eqnarray}
Next, we introduce the retarded Green function $\tilde{D}_{{\bf q}\lambda\lambda'}(t)$ in the usual manner:
\begin{eqnarray}
\tilde{D}_{{\bf q}\lambda\lambda'}(t) &=& -i\langle[\tilde{x}_{{\bf q}\lambda}(t),\tilde{x}_{-{\bf q}\lambda'}(0)]\rangle\theta(t),\\
        &\equiv& \frac{1}{2\pi}\int dz e^{-izt}\tilde{D}_{{\bf q}\lambda\lambda'}(z).
\end{eqnarray}
The equation of motion for $\tilde{D}_{{\bf q}\lambda\lambda'}(t)$ has the following form:
\begin{eqnarray}
\sum_{\lambda''}\Bigg[(-\frac{\partial^2}{\partial t^2}-\tilde{\omega}_{{\bf q}\lambda}^2)\delta_{\lambda\lambda''}-\frac{k^{o}_{\lambda''\lambda}({\bf q})+k^{o}_{\lambda\lambda''}(-{\bf q})}{2}\Bigg] \nonumber\\
\times\tilde{D}_{{\bf q}\lambda''\lambda'}(t) = \delta(t)\delta_{\lambda\lambda'}.
\end{eqnarray}
It is useful to employ $D_{{\bf q}\lambda\lambda'}(t)$, which is defined as
\begin{eqnarray}
D_{{\bf q}\lambda\lambda'}(t) \equiv \sqrt{2\tilde{\omega}_{{\bf q}\lambda}}\tilde{D}_{{\bf q}\lambda\lambda'}(t)\sqrt{2\tilde{\omega}_{{\bf q}\lambda'}}.
\end{eqnarray}
The Fourier components of $D_{{\bf q}\lambda\lambda'}(t)$ obey the following coupled equations:
\begin{eqnarray}
\sum_{\lambda''}\frac{1}{\sqrt{2\tilde{\omega}_{{\bf q}\lambda}}}\Bigg[(z^2-\tilde{\omega}_{{\bf q}\lambda}^2)\delta_{\lambda\lambda''}-\kappa_{\lambda\lambda''}({\bf q})\Bigg] \frac{1}{\sqrt{2\tilde{\omega}_{{\bf q}\lambda''}}}\nonumber\\
\times D_{{\bf q}\lambda''\lambda'}(z) = \delta_{\lambda\lambda'}.\label{GreenEQphonon}
\end{eqnarray}
Here, we have written the mixing term $\frac{1}{2}(k^{o}_{\lambda''\lambda}({\bf q})+k^{o}_{\lambda\lambda''}(-{\bf q}))$ as $\kappa_{\lambda\lambda''}({\bf q})$.
The energy eigenvalues are determined by the poles of $D_{{\bf q}\lambda\lambda}(z)$. It is evident that these poles coincide with those in usual basis, which diagonalize $k^{e}+k^{o}$.

If we take the electron-phonon coupling into account, we must introduce the polarization functions $\Pi_{{\bf q}\lambda\lambda''}(z)$ into the left-hand side of eq. (\ref{GreenEQphonon}). Defining $(D_{0{\bf q}}^{-1}(z))_{\lambda\lambda''}$ as the matrix in the large brackets of the left-hand side in eq. (\ref{GreenEQphonon}) divided by $\sqrt{2\tilde{\omega}_{{\bf q}\lambda}}\sqrt{2\tilde{\omega}_{{\bf q}\lambda''}}$, the equation of motion for $D_{{\bf q}\lambda''\lambda'}(z)$ in the interacting electron-phonon system can be written as
\begin{eqnarray}
\sum_{\lambda''}\Bigg[ (D_{0{\bf q}}^{-1}(z))_{\lambda\lambda''}-\Pi_{{\bf q}\lambda\lambda''}(z)\Bigg]D_{{\bf q}\lambda''\lambda'}(z) \!\!\!\!&=&\!\!\!\! \delta_{\lambda\lambda'}.\label{GreenEQphonon3}
\end{eqnarray}

\begin{figure}[b]
	\begin{center}
    \includegraphics[width=.5\textwidth]{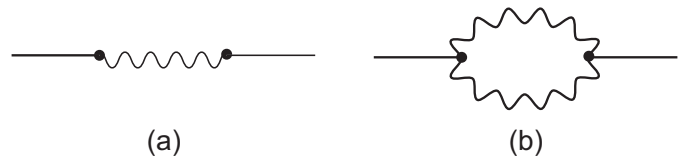}
  \end{center}
\caption{Diagram of the hybridization process between the acoustic and optical phonons near ${\bf q}=0$. The straight (wavy) line indicates the acoustic (optical) phonon propagator. (a) Harmonic term. (b) Anharmonic term.}
\label{fig-harmoanharmo}
\end{figure}

In order to clarify the importance of $\kappa_{\lambda\lambda''}({\bf q})$, it is instructive to consider the simplest case in which there exist one acoustic (its energy $\tilde{\omega}_{{\bf q}1}$) and one optical mode ($\tilde{\omega}_{{\bf q}2}$) in a system we consider. Furthermore, the electron-phonon couplings are nonzero only for the optical mode and $\kappa_{\lambda\lambda}({\bf q})=0$ by definition. Here, we use the words ``acoustic'' and ``optical'' with respect to those in the basis $\tilde{x}_{{\bf q}\lambda}$. In the diagonalized basis, there is no mixing between the acoustic and optical phonons in the harmonic Hamiltonian (\ref{Hpgeneral}). Then, eq. (\ref{GreenEQphonon3}) is reduced to
\begin{eqnarray}
\left(
\begin{array}{@{\,}cc@{\,}}
z^2-\tilde{\omega}_{{\bf q}1}^2 & - \kappa_{12}({\bf q}) \sqrt{ \frac{\tilde{\omega}_{{\bf q}1}}{\tilde{\omega}_{{\bf q}2}} } \\
   -\kappa_{21}({\bf q})\sqrt{\frac{\tilde{\omega}_{{\bf q}2}}{\tilde{\omega}_{{\bf q}1}}} & z^2-\tilde{\omega}_{{\bf q}2}^2-2\tilde{\omega}_{{\bf q}2}\Pi_{{\bf q}22}(z) \\
\end{array}
\right) \nonumber\\
\times\left(
\begin{array}{@{\,}cc@{\,}}
D_{{\bf q}11}(z) & D_{{\bf q}12}(z)\\
D_{{\bf q}21}(z) & D_{{\bf q}22}(z)
\end{array}
\right)=
\left(
\begin{array}{@{\,}cc@{\,}}
   2\tilde{\omega}_{{\bf q}1} & 0 \\
   0 & 2\tilde{\omega}_{{\bf q}2} \\
\end{array}
\right).\label{simpleD}
\end{eqnarray}
We find the form of the Green function for the acoustic mode as 
\begin{eqnarray}
2\tilde{\omega}_{{\bf q}1}D^{-1}_{{\bf q}11}(z)= z^2-\tilde{\omega}_{{\bf q}1}^2-\frac{|\kappa_{12}({\bf q})|^2}{z^2-\tilde{\omega}_{{\bf q}2}^2-2\tilde{\omega}_{{\bf q}2}\Pi_{{\bf q}22}(z)}.\label{Green11}
\end{eqnarray}
 Noted that in eq. (\ref{Green11}), $\kappa_{12}({\bf q}) \simeq \kappa_0(\hat{q}) |{\bf q}|$, $\tilde{\omega}_{{\bf q}2} \simeq\tilde{\Omega}=$ const. and $\tilde{\omega}_{{\bf q}1} \simeq \tilde{u}_0(\hat{q})|{\bf q}|$ for small ${\bf q}$, where $\hat{q}\equiv{\bf q}/|{\bf q}|$. Although the expression $\tilde{\omega}_{{\bf q}1} \simeq \tilde{u}_0(\hat{q})|{\bf q}|$ is not satisfied in general, it is a reasonable assumption for filled skutterudite and clathrate compounds. Diagrammatically, eq. (\ref{Green11}) is represented by Fig. \ref{fig-harmoanharmo}(a). Thus, the frequency-dependent phonon velocity $u(z,\hat{q})$ is given formally by
\begin{eqnarray}
u^2(z,\hat{q}) = \tilde{u}_0^2(\hat{q})+ \frac{|\kappa_0(\hat{q})|^2}{z^2-\tilde{\Omega}^2-2\tilde{\Omega}\Pi_{\hat{q}22}(z)}.\label{velo0}
\end{eqnarray}
In the limit of ultrasound, we can ignore $z\ll\tilde{\Omega}$ and set $\Pi_{\hat{q}22}=\Pi_{\hat{q}}^{\prime}-i\Gamma_{\hat{q}}z$ in eq. (\ref{velo0}), leading to
\begin{eqnarray}
u^2(z,\hat{q}) &=& \tilde{u}_0^2(\hat{q})- \frac{|\kappa_0(\hat{q})|^2}{\tilde{\Omega}^2+2\tilde{\Omega}\Pi_{\hat{q}}^{\prime}}\frac{1}{1-i2\Gamma_{\hat{q}} z/(\tilde{\Omega}+2\Pi_{\hat{q}}^{\prime})},\nonumber\\
&=& u^2(\infty,\hat{q})+\frac{u^2(0,\hat{q})-u^2(\infty,\hat{q})}{1-i\tau_{\hat{q}} z},\label{velo1}
\end{eqnarray}
where $\tau_{\hat{q}}\equiv \Gamma_{\hat{q}}/(\tilde{\Omega}+2\Pi_{\hat{q}}^{\prime})$. Equation (\ref{velo1}) corresponds to the phenomenological expression eq. (\ref{eq1}) used in the analysis of the ultrasound experiments. In phenomenological theory, the temperature dependence of $\tau_{\hat{q}}$ is assumed to be proportional to $\exp(E_{\Delta}/T)$. The activation scale $E_{\Delta}$ is expected to be of the order of $\tilde{\Omega}$. The fact that the ultrasound experimental results are well explained by eq. (\ref{eq1}) indicates the existence of a soft mode of the phonon (in this simple example, the soft mode, which appears around zero frequency, is related to the many-body effects between the optical phonon and conduction electrons). In the above simple model, we have ignored $\Pi_{{\bf q}nm}$ with $(n,m)\not = (2,2)$. This is because we consider that most of the important properties come from the self-energy of the optical phonon. If these come from the degrees of freedom in the lattice vibrations, it is clear that $E_{\Delta}$ cannot appear without the finite energy excitation of the optical phonon. 
 
\subsection{Anharmonic coupling}\label{anharmoniccoupling}
The discussion in the previous subsection, however, is not directly applicable to the rare-earth mode in filled skutterudites because the optical phonons that hybridize with the acoustic phonons near ${\bf q}=0$ are those with even parity in that system. The rare-earth mode couples with the acoustic phonon only through the anharmonic coupling so long as we ignore the electron-acoustic phonon interactions, as shown in Fig. \ref{fig-harmoanharmo}(b). Since the ultrasound generates strain $\varepsilon_{\Gamma}({\bf q})$ with the wave number ${\bf q}$ and the symmetry $\Gamma=\Gamma_1^{+},\ \Gamma_3^+$ and $\Gamma_5^+$ in the cubic $O_h$ point group, the rare-earth mode can couple with these strain-fields by constructing the direct product: $\Gamma_4^{-}\otimes \Gamma_4^{-}=\Gamma_1^{+}\oplus\Gamma_3^+\oplus\Gamma_5^+$. ($\varepsilon_{\Gamma}({\bf q})$ is expressed by the acoustic phonon fields.) Thus, we have coupling due to the anharmonicity as
\begin{eqnarray}
V^-=\sum_{\Gamma}g^-_{\Gamma}\varepsilon_{\Gamma}({\bf q})(\phi_-\phi_-)_{-{\bf q}\Gamma},
\end{eqnarray}
where $\phi_-$ is the phonon field with the odd parity and $g^-_{\Gamma}$ is the coupling constant. Here we use the symbolic notation $(\phi_-\phi_-)_{-{\bf q}\Gamma}$. For instance, $(\phi_-\phi_-)_{-{\bf q}\Gamma_{5xy}^{+}}$ should be read as the Fourier component of $(\phi_{ix}\phi_{iy})$ with the wave number $-{\bf q}$. Corresponding to $V^-$, the coupling between the optical and acoustic phonons in the previous subsection can be rewritten as
\begin{eqnarray}
V^+=\sum_{\Gamma}g^+_{\Gamma}\varepsilon_{\Gamma}({\bf q})\tilde{x}_{-{\bf q}\Gamma}.
\end{eqnarray}
The EC's are renormalized by the susceptibility of $(\phi_-\phi_-)_{\Gamma}$ (denoted as $\chi_{\phi\phi}^{\Gamma}(z,{\bf q})$). A diagrammatic expression is shown in Fig. \ref{fig-harmoanharmo}(b). Information of $\chi_{\phi\phi}^{\Gamma}$ is required in order to obtain an explicit form of the sound velocity. This is more difficult than in the case of the previous subsection. In phenomenological treatment, we can assume $\chi_{\phi\phi}^{\Gamma}\propto 1/(1-iz\tau_q)$, which is the same form as that in eq. (\ref{velo1}).
\\

We summarize this section as follows:
\begin{itemize}
	\item High-temperature region: the sound velocity is given by ``original'' sound velocity $\tilde{u}_0(\hat{q})$ subtracted by the contribution from the  ``non-interacting'' optical phonon $|\kappa_0(\hat{q})|^2/(\tilde{\Omega}^2+2\tilde{\Omega}\Pi_{\hat{q}}^{\prime})$ (see eq. (\ref{velo1})) because of the smallness of $z\tau_{\hat{q}}$. This corresponds to the sound velocity obtained by diagonalizing $k^e+k^o$ with renormalized $\tilde{\Omega}$. Thus, the sound velocity is observed as a smaller value than that at lower temperatures as explained below.
        \item Low-temperature region: the relaxation time $\tau_q$ becomes very large around $z=0$, and eventually $z\tau_{\hat{q}}$ exceeds $1$. Note that the relaxation time of the optical phonon at the original pole is much smaller than $\tau_{\hat{q}}$. As a result, the second term in eq. (\ref{velo1}) does not affect the sound velocity. Thus, we observe only the ``original'' sound velocity $\tilde{u}_0(\hat{q})$.

\end{itemize}
These aspects are similar to those discussed by Yamada for the critical fluctuations of lattice systems \cite{YamadaHandbook}. In these studies, the system is assumed to be located near the structural phase transition and the imaginary part of the polarization was phenomenologically introduced. In this paper, however, we will calculate the explicit temperature dependence of $\tau_{\hat{q}}$, namely $\propto \exp(E_{\Delta}/T)$, based on a relevant microscopic model. In the next section, we investigate a simple model with respect to the discussions in \S \ref{harmoniccoupling} for simplicity. However, in principle, it is straightforward to discussion the same topic based on \S \ref{anharmoniccoupling}. 

\section{Model and Calculation}\label{sec:micro}

In this section, we concentrate on a model in which there is only one electron band and one local Einstein phonon. To carry out the complete calculation, it is desirable to evaluate the properties of the model with two kinds of phonons (whose Green functions are formally given by eq. (\ref{simpleD})) and conduction electrons. However, in order to obtain a qualitative understanding, we restrict ourselves in one local phonon coupled with conduction electrons and approximate the frequency dependence of the sound velocity in \S \ref{soundvelo}. 

One of the simplest models is the Holstein model \cite{Holstein}, which is given as
\begin{eqnarray}
H&=&\sum_{\sigma {\bf ij}}t_{{\bf ij}}c_{{\bf i}\sigma}^{\dg}c_{{\bf j}\sigma}+{\rm h.c.}+\Omega_E\sum_{{\bf i}}b_{{\bf i}}^{\dg}b_{{\bf i}}\nonumber\\
&& + \sum_{{\bf i}\sigma}g\Big(c_{{\bf i}\sigma}^{\dg}c_{{\bf i}\sigma}-\frac{1}{2}\Big)(b_{{\bf i}}^{\dg}+b_{{\bf i}}),\label{HolH}
\end{eqnarray}
where $t_{{\bf ij}}$ represents the hopping of conduction electrons, $c_{{\bf i}\sigma}^{\dg}$ is the creation operator of electrons at site ${\bf i}$ and spin $\sigma$, $b_{{\bf i}}^{\dg}$ is the phonon creation operator at site ${\bf i}$ with Einstein energy $\Omega_E$, and $g$ is the electron-phonon coupling constant. This model has been discussed for the past five decades\cite{EngelsbergSch,Freericks,Capone}.  Recently, the phase diagram of this model at $T=0$ was discussed based on the dynamical mean field theory (DMFT) \cite{DMFT} using the numerical renormalization group (NRG) as an impurity solver \cite{MayerHews1, MayerHews2, Jeon1, Jeon2}. The phase diagram of a somewhat different model \cite{MitsumotoOno1} was discussed by DMFT, but with exact diagonalization. For finite temperatures, data is lacking for frequencies smaller than the temperature in the NRG method. Recent developments in calculating spectral functions in NRG enable us to obtain the spectral functions roughly in the range of $z \simge T\times 1/10$ \cite{FullFockNRG1,FullFockNRG2}. However, it is insufficient to discuss the soft mode at finite temperatures. Thus, in the following subsections, we introduce a simple self-consistent treatment of the phonon system, which can capture the essential aspect of the emergence of the soft mode at low temperatures. For the connection with the single site approximation, which is widely used for experimental analysis of the quadrupolar degrees of freedom in f-electron systems\cite{TalLuthi}, we restrict ourselves to the single-site problem of a local phonon interacting with conduction electrons. This model is called the Holstein-Anderson model, which will be explained below. 

\subsection{Self-consistent theory in pseudo-fermion representation of phonon}
First, we explain a pseudo-fermion mapping of phonon operators. The displacement operator $\hat{X}=b^{\dg}+b$ has the matrix elements in the phonon Hilbert space as follows:
\begin{eqnarray}
X_{nm}&\equiv&\langle n| \hat{X} | m\rangle\nonumber\\
&=& \sqrt{m}\delta_{m,n+1}+\sqrt{n}\delta_{m,n-1},
\end{eqnarray}
where $n$ and $m$ are eigenvalues of the number of the phonons ($b^{\dagger}b$). We introduce pseudo-fermions $a_n^{\dag}$, which create the state with phonon numbers $n=0,1,2,\cdots$, as reported by Abrikosov for the case of ``spin'' in the Kondo problem \cite{AbriPseudo}. Using the pseudo-fermions, the displacement operator is represented by
\begin{eqnarray}
  \hat{X}=\sum_{n,m}a^{\dag}_nX_{nm}a_{m}.
\end{eqnarray}

In this representation, the Holstein-Anderson model with the Coulomb repulsion of the electrons $U=0$ becomes
\begin{eqnarray}
H\!\!\!\!\!&=&\!\!\!\!\!\sum_{{\bf k}\sigma}[{\epsilon}_{\bf k}{c}_{{\bf k}\sigma}^{\dg}{c}_{{\bf k}\sigma}+(v c_{{\bf k}\sigma}c^{\dag}_{0\sigma}+{\rm h.c.})]+\sum_{n}n\Omega_Ea_{n}^{\dg}a_{n} \nonumber\\
\!\!\!\!\!&+&\!\!\!\!\! g\sum_{\sigma}(c_{0\sigma}^{\dg}c_{0\sigma}-\frac{1}{2})a_{n}^{\dag}X_{nm}a_{m}+\sum_{n}\lambda(a_{n}^{\dag}a_{n}-1), \label{HolAnderson}
\end{eqnarray}
where $c_{0\sigma}^{\dag} (a_{n}^{\dg})$ is the on-site conduction electron (pseudo-fermion) creation operator with spin $\sigma$, ${\epsilon}_{\bf k}$ and $v$ characterize the conduction electron dispersion and the hybridization, respectively, and ${c}_{{\bf k}\sigma}^{\dg}$ is the creation operator of the conduction electron with the wave vector ${\bf k}$ and spin $\sigma$. We have introduced a Legendre multiplier $\lambda$ in order to prohibit double occupancy of the pseudo-fermions.

Next, we explain a self-consistent treatment of this impurity model. The method explained here is similar to the self-consistent ladder approximation in the Coqblin-Schrieffer model with crystalline-electric-field states of Ce impurities \cite{Maekawa, kashiba} and non-crossing approximation (NCA) of Anderson model \cite{NCA2,Bickers}. We define the Matsubara Green functions of $a_n$ and $c_{0\sigma}$ in the imaginary time $\tau$ as follows:
\begin{eqnarray}
A_{nm}(\tau)&=&-\langle T_{\tau}a_n(\tau)a_m^{\dag}(0)\rangle,\\
G(\tau)&=&-\langle T_{\tau}c_{0\sigma}(\tau)c_{0\sigma}^{\dag}(0)\rangle,
\end{eqnarray}
where $T_{\tau}$ is the time-order operator, and we omit the spin dependence in $G(\tau)$ hereafter.

\begin{figure}[t!]
	\begin{center}
    \includegraphics[width=0.5\textwidth]{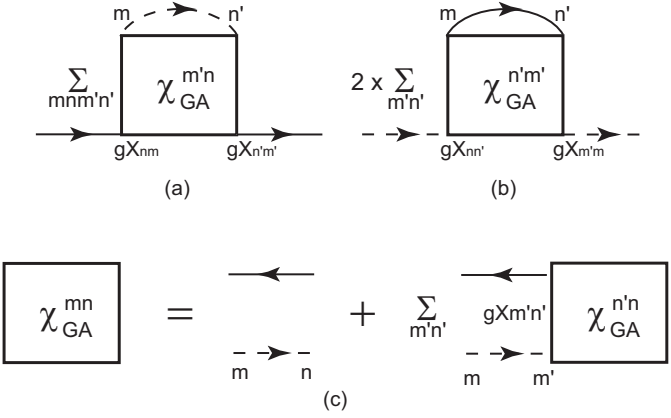}
  \end{center}
\caption{Self-energy diagrams (a) $\Sigma_G(i\omega_n)$ and (b) $\Sigma_A^{nm}(i\omega_n)$. (c) Diagrammatic equation for $\chi_{GA}$. The solid (dashed) line represents $G(A)$.}
\label{fig-3-7}
\end{figure}

The non-interacting Green functions $A_{nm}^0(i\omega_n)$ and $G^0(i\omega_n)$, are given as
\begin{eqnarray}
A_{nm}^0(i\omega_n) &=& [i\omega_n-m\Omega_E-\lambda]^{-1}\delta_{nm},\\
G^0(i\omega_n) &=& [i\omega_n+\mu - \Delta(i\omega_n)]^{-1},
\end{eqnarray}
where
\begin{eqnarray}
\Delta(i\omega_n)&=&\sum_k\frac{|v|^2}{i\omega_n-\tilde{\epsilon_k}},\ \ \ \ \ \ \ \ \ \ \ 
\end{eqnarray}
where $\omega_n$ is the fermionic Matsubara frequency $\omega_n=(2n+1)\pi T$,
In order to take into account the electron-phonon coupling, we consider the self-energy diagrams $\Sigma_A^{nm}(i\omega_n)$ and $\Sigma_G(i\omega)$ shown in Figs. \ref{fig-3-7}(a) and \ref{fig-3-7}(b), respectively, as:
\begin{eqnarray}
\Sigma_A^{nm}(i\omega_n)&=&2g^2T\sum_{i\epsilon_m}G(i\epsilon_m)\nonumber\\
&\times&\sum_{n'm'}X_{nn'}\chi_{GA}^{n'm'}(i\omega_n-i\epsilon_m)X_{m'm},\\
\Sigma_G(i\omega_n)&=&g^2T\sum_{i\epsilon_m}\sum_{nmn'm'}A_{m'n}(i\epsilon_m)\nonumber\\
&\times&\chi_{GA}^{mn'}(i\epsilon_m-i\omega_n)X_{nm}X_{n'm'}.\label{MatsuSigmaG}
\end{eqnarray}
Here, $\chi_{GA}$ is the solution for the diagrammatic equation given in Fig. \ref{fig-3-7}(c), which is reduced to the summation of the infinite series of the ladder diagram $\chi_{GA}^0$ constituted of $G$ and $A$. This is analytically given by
\begin{eqnarray}
\sum_{n'}[\delta_{nn'}-g\sum_{m'}{\chi_{GA}^0}^{nm'}(i\nu_l)X_{m'n'}]\chi_{GA}^{n'm}(i\nu_l)\nonumber\\
={\chi_{GA}^0}^{nm}(i\nu_l), \label{ladd}
\end{eqnarray}
where $\nu_l=2\pi l T$ is a bosonic Matsubara frequency.
In matrix form, eq. (\ref{ladd}) is expressed as
\begin{eqnarray}
 \hat{\chi}_{GA}=[{\bf{1}}-g\hat{\chi}_{GA}^0\hat{X}]^{-1}\hat{\chi}_{GA}^0. \label{laddMAT}
\end{eqnarray}
The explicit form of $\chi_{GA}^0$ is given by
\begin{eqnarray}
{\chi_{GA}^0}^{nm}(i\nu_l)=-T\sum_{i\epsilon_m}G(i\epsilon_m)A_{nm}(i\nu_l+i
\epsilon_m).
\end{eqnarray}
Note that the first-order term in $g$ in $\Sigma_A$ and $\Sigma_G$ vanishes, together with the term $\frac{1}{2}$ in the interaction term in eq. (\ref{HolAnderson}).

Using $\Sigma_A^{nm}$ and $\Sigma_G$ together with Dyson's equation: 
\begin{eqnarray}
A_{nm}(i\omega_n)\!\!\!\!&=&\!\!\!\!A^0_{nm}(i\omega_n)\nonumber\\&+&\!\!\!\!\!\!\sum_{n'm'}A^0_{nn'}(i\omega_n)\Sigma^{n'm'}_A(i\omega_n)A_{m'm}(i\omega_n),\label{DyeqA}\\
G(i\omega_n)\!\!\!\!&=&\!\!\!\!G^0(i\omega_n)+G^0(i\omega_n)\Sigma_G(i\omega_n)G(i\omega_n),
\end{eqnarray}
we can obtain the self-consistent Green functions. In actual calculations, it is  better to treat the retarded Green functions $\Sigma_A^{nm}(z+i0)$ and $\Sigma_G(z+i0)$. Carrying out the analytic continuations, we can obtain the set of equations for $\Sigma_A^{nm}(z+i0)$ and $\Sigma_G(z+i0)$, as shown in the Appendix (we hereinafter write $z+i0$ as simply $z$).  

The phonon Green function $D(\tau)$, which is the most crucial quantity in the present paper, is given by
\begin{eqnarray}
D(\tau)\!\!\!\!\!&=&\!\!\!\!\!-\langle T_{\tau} \hat{X}(\tau)\hat{X}(0)\rangle\nonumber\\
       \!\!\!\!\!&=&\!\!\!\!\!-\sum_{nn'mm'}X_{nm}X_{n'm'}\langle T_{\tau} a^{\dag}_n(\tau)a_{m}(\tau)a^{\dag}_{n'}(0)a_{m'}(0)\rangle\nonumber\\
       \!\!\!\!\!&\simeq&\!\!\!\!\sum_{nn'mm'}X_{nm}X_{n'm'}A_{mn'}(\tau)A_{m'n}(-\tau),\label{Dfunc}
\end{eqnarray}
where we ignore the vertex corrections in the last line. In this approximation, the phonon Green function in Matsubara frequency $\nu_l$ is given by
\begin{eqnarray}
D(i\nu_l)\!\!\!\!&=&\!\!\!\!\!\!\!\!\!\sum_{nn'mm'}\!\!\!\!\!X_{nm}X_{n'm'}T\sum_{i\omega_{l'}}A_{mn'}(i\omega_{l'}+i\nu_l)A_{m'n}(i\omega_{l'}).\nonumber\\
\end{eqnarray}

The main feature of the approximation scheme explained in this subsection is to take into account ``phonon fluctuations''. As such, the results obtained in this method might overestimate the fluctuations because we ignore the vertex corrections in eq. (\ref{Dfunc}). Despite this, the present approximation can visualize the essential points of the low-energy phenomena in the present electron-phonon system.

\subsection{Numerical results}
In this subsection, we show the numerical results of the Holstein-Anderson model (\ref{HolAnderson}). For the conduction electron, we assume the Gaussian density of states $\rho(z)$:
\begin{eqnarray}
\rho(z)=\frac{1}{\sqrt{2\pi t^2}}\exp(-\frac{z^2}{2t^2}),
\end{eqnarray}
which corresponds to an infinite dimensional hyper-cubic lattice with the nearest neighbor hopping $t$. We set $t=0.2$ and $\Omega_E=0.01$ throughout this subsection. The hybridization width Im$\Delta(z)$ is given by using $\rho(z)$ as
\begin{eqnarray}
  -{\rm Im}\Delta(z)=\pi v^2\rho(z),
\end{eqnarray}
where $v$ is a constant. The real part of $\Delta(z)$ is calculated by the Kramers-Kronig relation
\begin{eqnarray}
  {\rm Re}\Delta(z)={\rm P}\int\frac{dx}{\pi}\frac{{\rm Im}\Delta(x)}{x-z},
\end{eqnarray}
where the integral is taken as its principle value. In the numerical calculations below, we introduce the cutoff number for the pseudo-fermion: $N_{\rm cut}$, {\it i.e.,} $a_0,\ a_1,\ \cdots,\ a_{N_{\rm cut}-1}$. Due to this restriction, we can calculate the dynamical quantities only in $T \simle (N_{\rm cut}-1)\Omega_E$. However, as shown below, this restriction does not become serious in the temperature regions of interest.
\begin{figure}[t!]
	\begin{center}
    \includegraphics[width=0.5\textwidth]{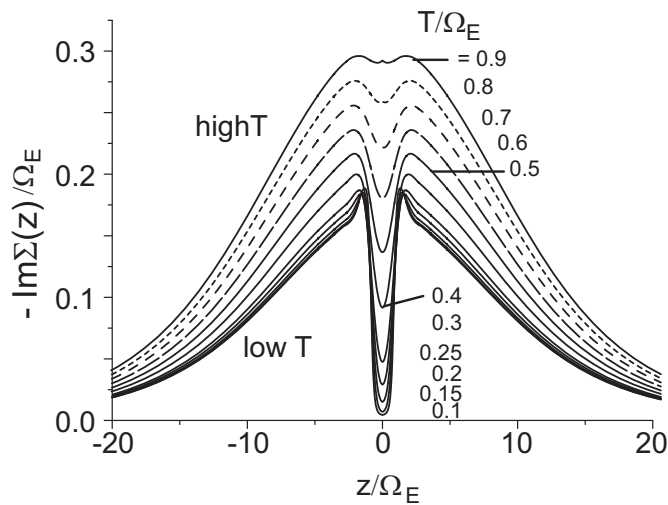}
  \end{center}
\caption{Imaginary part of the self-energy of the electron Im$\Sigma_G$ with $N_{\rm cut}=6$, $v=0.1$, and $g=0.01$.}
\label{fig-ImSig}
\end{figure}
\begin{figure}[h!]
	\begin{center}
    \includegraphics[width=0.5\textwidth]{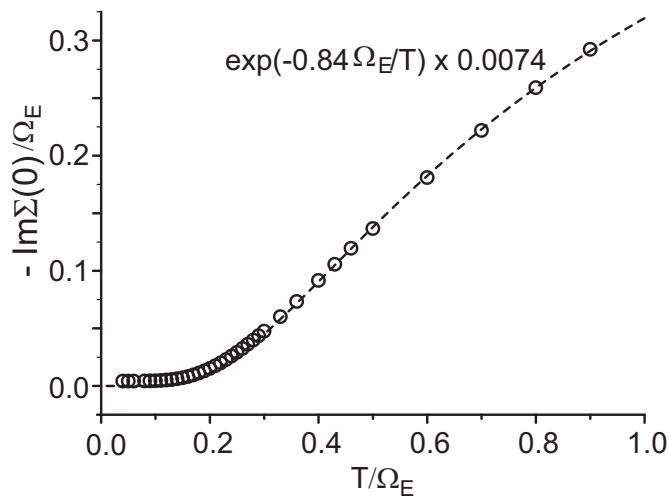}
  \end{center}
\caption{Temperature dependence of Im$\Sigma_G(0)$ from Fig. {\ref{fig-ImSig}}. The dashed line is the curve of $0.0074\times\exp(-0.84\Omega_E/T)$.}
\label{fig-TdepImSig0}
\end{figure}

\vspace{3mm}
{\raggedright
{\it - Electron self-energy -}}
\vspace{3mm}

 We show the result of the imaginary part of the electron self-energy Im$\Sigma_G$ in Fig. \ref{fig-ImSig}. As the temperatures decrease, a dip structure is developed around $|z|<\Omega_E$. This simply represents the fact that an electron with an energy below $\Omega_E$ cannot emit the Einstein phonon with energy $\Omega_E$. Thus, the temperature dependence of Im$\Sigma_G(z)$ for $z\sim 0$ becomes the activation-type, $\propto \exp(\alpha/T)$, where $\alpha$ is a constant on the order of $\Omega_E$, as shown in Fig. \ref{fig-TdepImSig0}.

\begin{figure}[t!]
	\begin{center}
\includegraphics[width=0.5\textwidth]{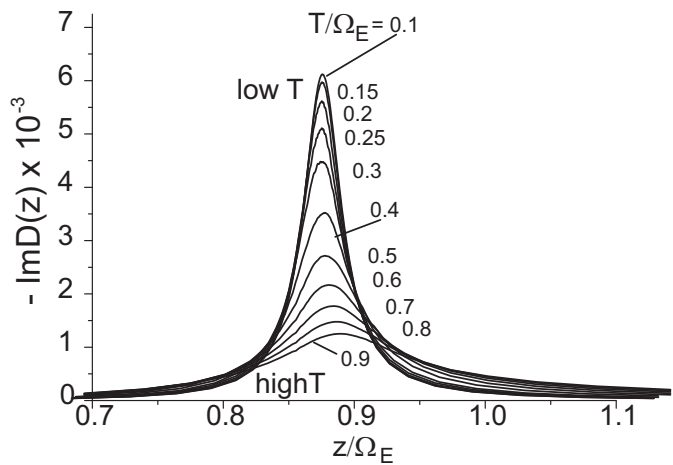}
\end{center}
\caption{Frequency dependence of the phonon Green function at high energy. The parameters are $g=0.01$, $N_{\rm cut}=6$, and $v=0.1$.}
\label{fig-phononHIGH}
\end{figure}
\begin{figure}[t!]
	\begin{center}
\includegraphics[width=0.5\textwidth]{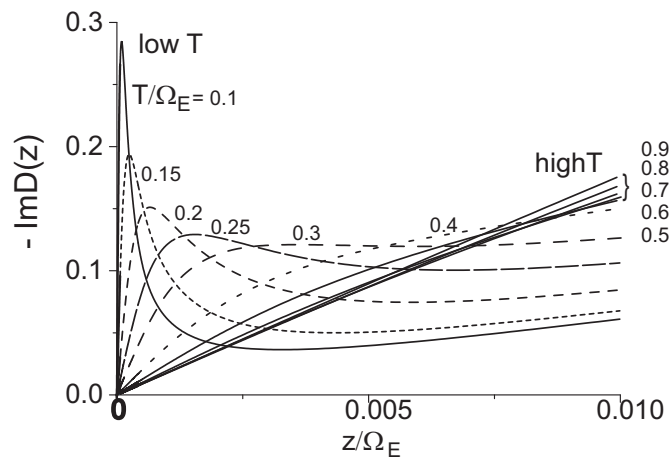}
\end{center}
\caption{Low-frequency phonon Green function with same parameters as in Fig. \ref{fig-phononHIGH}. }
\label{fig-phononLOW}
\end{figure}

\vspace{3mm}
{\raggedright
{\it - Phonon Green functions -}}
\vspace{3mm}

\begin{figure}[t!]
	\begin{center}
    \includegraphics[width=0.5\textwidth]{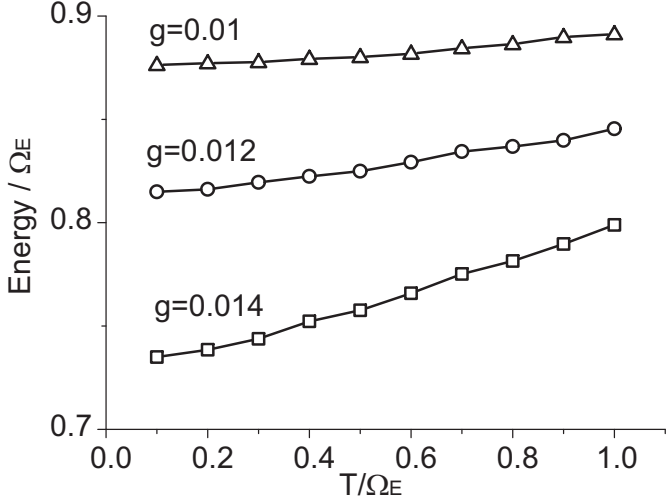}
  \end{center}
\caption{Temperature dependence of the high-energy peak in Im$D(z)$ with different values of $g$. Other parameters are as in Fig. \ref{fig-phononHIGH}.}
\label{fig-TdepMainPeak}
\end{figure}

In Figs. \ref{fig-phononHIGH} and \ref{fig-phononLOW}, we show the frequency and temperature dependence of the phonon Green function $-{\rm Im}D(z)$ in high- (Fig. \ref{fig-phononHIGH}) and low- (Fig. \ref{fig-phononLOW}) frequency regions. The position of the main peak around $\Omega_E$ decreases as temperatures decrease, as shown in Fig. \ref{fig-TdepMainPeak}. The peak width also decreases at low temperatures, making the peak sharper. This main peak corresponds to the excitations $a_0\leftrightarrow a_1,\ a_1 \leftrightarrow a_2,\cdots$ in the pseudo-fermion picture. At low temperatures, below $T\sim 0.3\Omega_E$, a low-lying excitation becomes prominent, as shown in Fig. \ref{fig-phononLOW}. This low-energy soft mode arises from the development in the off-diagonal elements of pseudo-fermion Green functions $\bar{A}_{01}(0),\ \bar{A}_{12}(0),\ \cdots$, and $\bar{A}_{nn}(0)$ for $n\ne 0$.

\begin{figure}[t!]
	\begin{center}
\includegraphics[width=0.5\textwidth]{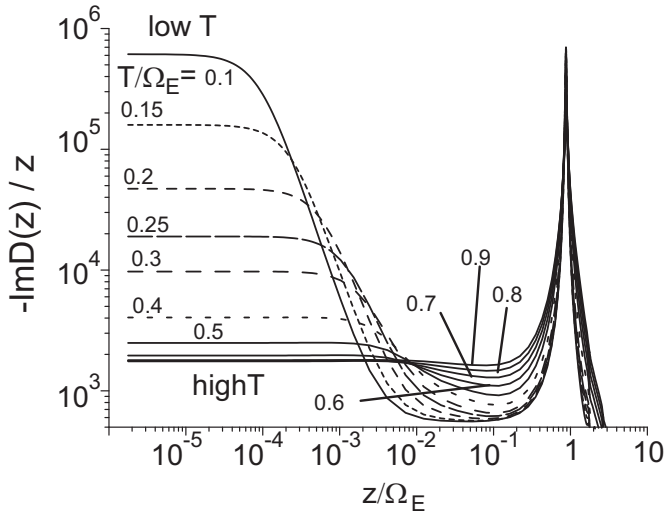}
\end{center}
\caption{Frequency dependence of $-{\rm Im}D(z)/z$ on a log-log scale with different temperatures. The parameters are as in Fig. \ref{fig-phononHIGH}}
\label{fig-spectraPHONON}
\end{figure}
\begin{figure}[h!]
	\begin{center}
    \includegraphics[width=0.5\textwidth]{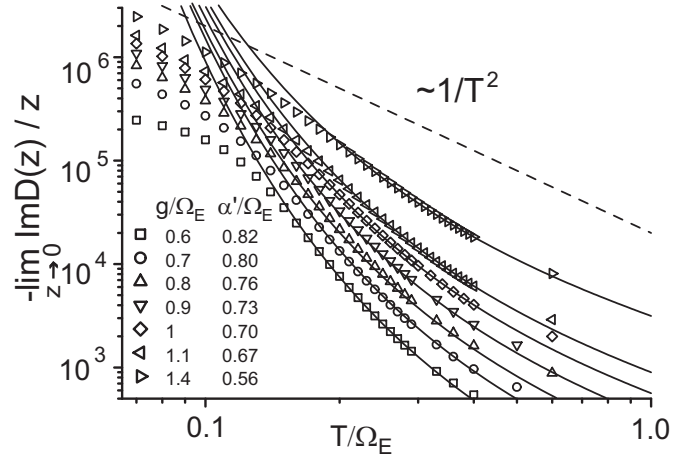}
\end{center}
\caption{Temperature dependence of $-\lim_{z\to 0}{\rm Im}D(z)/z$ for various $g$. The lines are curves of $\propto\exp(\alpha'/T)/T$. The other parameters are as in Fig. \ref{fig-phononHIGH}}
\label{fig-Temp-slop}
\end{figure}

Note that the slope at the origin maintains its linear frequency dependence, {\it i.e.,} $-{\rm Im}D(z)/z=$ const. near $z=0$, as shown in Fig. \ref{fig-spectraPHONON}. Thus, this peak is not generated by fictitious increases of the dynamical susceptibility ${\rm Im}\chi(z)/z$ that frequently arise in NCA at low temperatures \cite{KojimaKuramoto, Otsuki}. The temperature dependence of this slope is shown in Fig. \ref{fig-Temp-slop}. At intermediate temperatures ($0.1\simle T/\Omega_E\simle 1$), the temperature dependence of the slope is well fitted by the activation-type function, $\exp(\alpha'/T)/T$, where $\alpha'$ is a constant on the order of $\Omega_E$. At low temperatures $(T/\Omega_E\simle 0.1)$, the data deviates from the exponential dependence and exhibits $1/T^2$ behavior. These behaviors are discussed below.

In order to clarify the origin of the temperature dependence of Im$D(z)$, it is useful to consider Im$\bar{\Sigma}_A^{nm}(0)$. In the lowest order ladder diagram of $\chi_{GA}$, Im$\bar{\Sigma}_A^{nm}$ is written as
\begin{eqnarray}
{\rm Im}\bar{\Sigma}_A^{nm}(0)&=&2g^2\int\frac{dy}{\pi}n_{B}(y){\rm Im}\chi_G(y)\nonumber\\&\times&\sum_{n'm'}X_{nn'}{\rm Im}\bar{A}_{n'm'}(y)X_{m'm}.\label{explainT}
\end{eqnarray}
Here, $n_B$ is the Bose function and Im$\chi_{G}$ is given by 
\begin{eqnarray}
{\rm Im}\chi_G(z)&=&-\int\frac{dy}{\pi}[n_F(y+z)-n_F(y)]\nonumber\\
&\times&{\rm Im}G(y+z){\rm Im}G(y).
\end{eqnarray}
At low temperatures ($T\simle0.1\Omega_E$ in Fig. \ref{fig-Temp-slop}, referred to herein as region I), taking into account the fact that the temperature dependence in $\lim_{z\to0}$Im$D(z)/z$ comes from the contribution of the low-energy peak in Im$\bar{A}_{nm}$, we can replace Im$\bar{A}_{n'm'}$ in eq. (\ref{explainT}) by $\propto\delta(y)$. This yields Im$\bar{\Sigma}^{nm}_A(0)\propto T$. On the other hand, at higher temperatures ($0.1\Omega_E\simle T\simle\Omega_E$ in Fig. \ref{fig-Temp-slop}, referred to herein as region II), the strength of Im$\bar{A}_{nm}(0)$ is weak and the temperature dependence comes from $\sim n_B(\Omega_E^{\prime})$Im$\bar{A}_{n'm'}(\Omega_E^{\prime})$, where $\Omega_E^{\prime}$ is the renormalized Einstein energy. Although $\Omega_E^{\prime}$ is weakly temperature dependent, for simplicity, we regard $\Omega^{\prime}_E$ as a temperature independent parameter. These observations yield Im$\bar{\Sigma}^{nm}_A(0)\propto n_B(\Omega^{\prime}_E)\simeq\exp(-\Omega^{\prime}_E/T)$, for $T<\Omega^{\prime}_E$. 

For the phonon Green function, eq. (\ref{ImD}) is reduced, at low temperature and frequency with $z\ll T$ to
\begin{eqnarray}
{\rm Im}D(z)&\simeq&-\frac{z}{T}\int \frac{dy}{\pi}e^{-y/T}\sum_{nn'mm'}X_{nm}X_{n'm'}\nonumber\\ &\times& {\rm Im}\bar{A}_{mn'}(y+z){\rm Im}\bar{A}_{m'n}(y),\label{ImD2}
\end{eqnarray}
where we have used $\Big[ -\sum_l\int \frac{dx}{\pi}e^{-x/T}{\rm Im}\bar{A}_{ll}(x) \Big]\simeq 1$ for $T\simle\Omega_E$. To simplify eq. (\ref{ImD2}) further, we omit the indices $n,\ m,\ n',\ {\rm and}\ m'$. Taking into account the fact that the dominant contribution comes from $y=0$ in the integrand, and $-{\rm Im}\bar{A}(y)\sim (\gamma/2)/(y^2+(\gamma/2)^2)$ with $\gamma=-2{\rm Im}\bar{\Sigma}_A(0)$ (this is of course too simplified an approximation), we can estimate $-{\rm Im}D(z)/z$ as
\begin{eqnarray}
-\frac{{\rm Im}D(z)}{z}&\propto&\frac{1}{T}\frac{\gamma}{z^2+\gamma^2}\stackrel{{z\to0}}{\longrightarrow}\frac{1}{T\gamma},\label{Debyeform}\\
  &\propto&\!\!\!\! \left\{
   \begin{array}{cl}
    T^{-2} & \mbox{ in I,}\\
    T^{-1}\exp(\Omega^{\prime}_E/T) & \mbox{ in II.}
   \end{array}
\right.\label{ImD3}
\end{eqnarray}
These estimations reproduce the numerical results shown in Fig. \ref{fig-Temp-slop}. The expression eq. (\ref{Debyeform}) is expected to be a relevant origin for the Debye-type formula in eq. (\ref{eq1}). {\it Indeed, the form of} eq. (\ref{Debyeform}) {\it is as same as an expression in the case in which there are degenerate modes with the lifetime $\tau=\gamma^{-1}$ in the ground state}\cite{Young}. This is manifest by the $1/T$ in front of eq. (\ref{Debyeform}).

\subsection{Frequency dependence on sound velocity}\label{soundvelo}
In order to estimate the frequency dependence on the sound velocity correction with the use of the Holstein-Anderson model, we note that the high-energy spectra Im$D(z)$ does not contribute to the frequency dependence in Re$D(z)$ for $z\ll \Omega_E$ through the Kramers-Kronig transformation. The high-frequency part only gives weak monotonic temperature dependence. Thus, we define Re$D_{\rm low}(z)$ by introducing the cutoff parameter $\Lambda<\Omega_E$ as:
\begin{eqnarray}
{\rm Re}D_{\rm low}(z)\equiv {\rm P}\int_{-\Lambda}^{\Lambda}\frac{dx}{\pi}\frac{{\rm Im}D(x)}{x-z}. \label{cutLambda}
\end{eqnarray}
From the discussions in \S \ref{sec:phenomenolo}, we can estimate the effect of the optical phonon on the sound velocity. We approximately replace the second term in eq. (\ref{velo0}) by the phonon Green function obtained in the previous single-site problem:
\begin{eqnarray}
u^2(z,\hat{q}) \simeq \tilde{u}_0^2(\hat{q})+ \frac{|\kappa_0(\hat{q})|^2}{2\tilde{\Omega}}D(z).\label{veloApp}
\end{eqnarray}
Although the present results cannot be directly compared to the experimental data in the case in which the parity of the optical mode is $-1$, it is expected that their qualitative behaviors do not differ greatly. 
\begin{figure}[t!]
	\begin{center}
    \includegraphics[width=.5\textwidth]{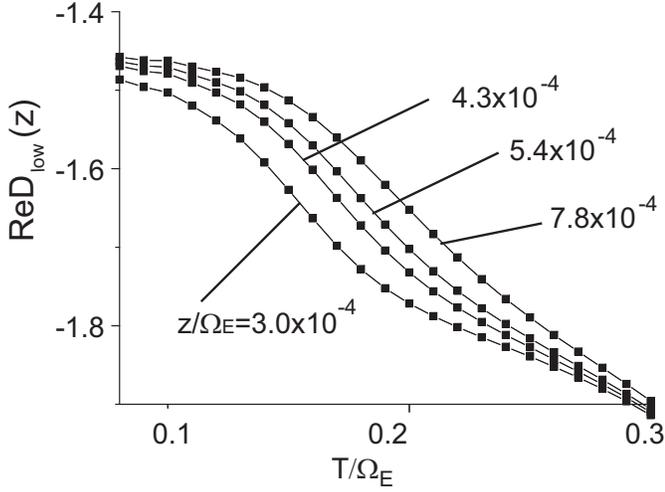}
\end{center}
\caption{Temperature dependence of ${\rm Re}D_{\rm low}(z)$ with different frequencies with $\Lambda=0.35\Omega_E=0.0035$. The parameters are as in Fig. \ref{fig-phononHIGH}: $g=0.01$, $N_{\rm cut}=6$ and $v=0.1$.}
\label{fig-Temp-Elas}
\end{figure}

In Fig. \ref{fig-Temp-Elas}, we show the temperature dependence of Re$D_{\rm low}(z)$ with different frequencies with $\Lambda=0.0035=0.35\times\Omega_E$ and $g=\Omega_E$. Although the choice of $\Lambda$ includes a certain arbitrariness, the essential feature of Re$D_{\rm low}$ does not change if we take $\Lambda$ to be smaller than $\simeq 0.5\Omega_E$. In the case of harmonic coupling, Re$D_{\rm low}(z)$ is directly related to the sound velocity and the EC (we cannot discuss the anisotropy in EC's, see \S \ref{Discussion}). For this case, the frequency-dependent term $C(z)$ in the EC's is simply given by

\begin{eqnarray}
C(z)\propto {\rm Re}D_{\rm low}(z).
\end{eqnarray}
 Clear hardenings, the positions of which shift to higher temperatures as the frequency increases, can be observed in Fig. \ref{fig-Temp-Elas}. This feature is consistent with the phenomenological expression (\ref{eq1}). 

\begin{figure}[t!]
	\begin{center}
    \includegraphics[width=0.45\textwidth]{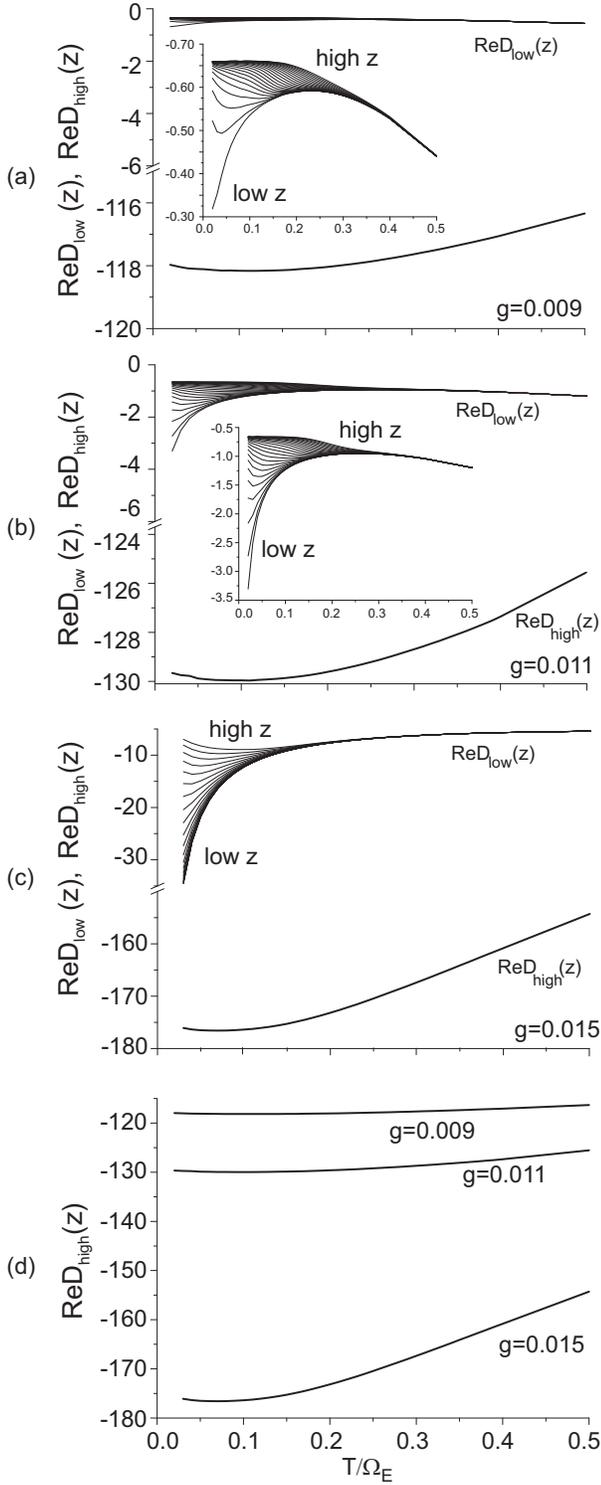}
\end{center}
\caption{(a)-(c) Temperature dependences of ReD$_{\rm low}$ and ReD$_{\rm high}$ at various frequencies $z=4.8\times 10^{-5}\Omega_E-1.1\times 10^{-3}\Omega_E$. (a) $g=0.009$, (b) $g=0.011$, and (c) $g=0.015$. The insets in (a) and (b) are close up views of ReD$_{\rm low}(z)$. ReD$_{\rm low}(z)$ is estimated by setting $\Lambda=0.25\Omega_E$ in eq. (\ref{cutLambda}). Other parameters are as in Fig. \ref{fig-phononHIGH}. (d) Temperature dependence of ReD$_{\rm high}$ at different $g$ taken from (a)-(c).}
\label{fig-total_FULL}
\end{figure}

In addition to Re$D_{\rm low}$, there are contributions from the high-energy spectra: Re$D_{\rm high}(z)\equiv$ Re$D(z)-$Re$D_{\rm low}(z)$ and the anharmonicity of acoustic phonons, both of which are essentially frequency independent. The elastic constant observed in the experiments is basically given by these two contributions unless we take into account electric contributions. The former makes the elastic constant soften at low temperatures and low frequency. On the other hand, the latter makes the elastic constant harden. We show the temperature, $g$ (coupling constant), and frequency dependence of Re$D_{\rm low}$ and Re$D_{\rm high}$ in Figs. \ref{fig-total_FULL}(a)-\ref{fig-total_FULL}(d). The magnitude of Re$D_{\rm low}$ increases as $g$ increases compared to the Re$D_{\rm high}$. Although, in Fig. \ref{fig-total_FULL}(a), there exist shoulders as a function of temperatures in Re$D_{\rm low}(z)$, as shown in the inset of Fig. \ref{fig-total_FULL}(a), their magnitudes are quite small relative to Re$D_{\rm high}$. On the other hand, these features disappear at larger $g$, as shown in Fig. \ref{fig-total_FULL}(c). The shoulder in the parameters in Fig. \ref{fig-total_FULL}(c) can be seen at the higher frequency (not shown in Fig. \ref{fig-total_FULL}(c)). In all of the data in Figs. \ref{fig-total_FULL}(a)- \ref{fig-total_FULL}(c), softening behaviors are observed at Re$D_{\rm low}(z)$ with small $z$. These features originate from the development of the soft mode being too strong in our calculation. More reliable methods are needed for the low-temperature region in order to discuss this point further.

\begin{figure}[t!]
	\begin{center}
    \includegraphics[width=0.5\textwidth]{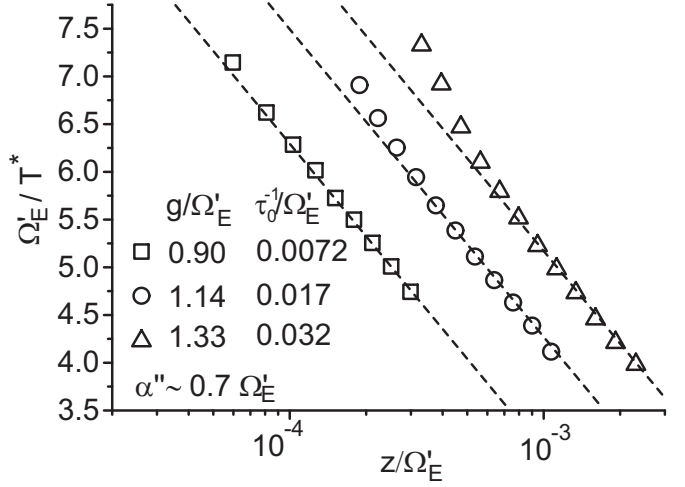}
\end{center}
\caption{Frequency dependence of $T^*$ for three different $g$ values, maintaining the renormalized Einstein frequency $\Omega_E^{\prime}$ constant. $\Omega_E^{\prime}$ is determined by the peak position of Im$D(z)$ at the high-energy region in the zero temperature limit. We plot the dashed lines, which are determined by the condition $z\tau_0\exp(\alpha''/T)=1$. The values of $\tau_0$ and $\alpha''$ are listed in the panel. The other parameters are as in Fig. \ref{fig-phononHIGH}. In particular, the data for $g/\Omega_E^{\prime}=1.14$ is the same as in Fig. \ref{fig-phononHIGH}.}
\label{fig-3-resonanceposition}
\end{figure}

In order to observe the temperature dependence of the shoulder in Re$D(z)$, it is useful to differentiate Re$D_{\rm low}(z)$ by $T$. In Fig. \ref{fig-3-resonanceposition}, we plot the temperatures $T^*$, where $|d{\rm Re}D_{\rm low}(z)/dT|$ takes a maximum value. Roughly speaking, $T^*$ corresponds to the midpoint of the shoulder. In Fig. \ref{fig-3-resonanceposition}, we show the results for three different parameter sets, keeping the renormalized Einstein frequency $\Omega_E^{\prime}$ the same for each set. As shown in Fig. \ref{fig-3-resonanceposition}, the frequency dependence of $\Omega_E^{\prime}/T^*$ is well described by the condition $z\tau=1$ with $\tau=\tau_0\exp(\alpha''/T)$. Here, as before, $\alpha''\simeq0.7\Omega_E^{\prime}$ is a parameter on the order of $\Omega_E^{\prime}$. Interestingly, the estimated values of $\tau_0^{-1}$ are approximately $\alpha''\times1/100$. This is in good qualitative agreement with the experimental values $\tau_0^{-1}=0.006\alpha''-0.1\alpha''$ for filled skutterudites and clathrate compounds. We can find that $\tau_0^{-1}$ becomes small as $g$ decreases. This arises from the fact that the soft mode amplitude becomes smaller and its frequency region moves to the lower side as the electron-phonon interaction $g$ becomes small. This feature is sensitive to the activation energy $\alpha''$. A small change in $\alpha''$ can affect the magnitude of $\tau_0$. This is the reason why we show the data with the same $\Omega_E^{\prime}$ values in Fig. \ref{fig-3-resonanceposition}.

\section{Discussions}\label{Discussion}
Thus far, we have restricted our study to the examination of a system with one component phonon. Here, we discuss qualitatively the anisotropy in the UD observed in the above experiments. 

In La$_3$Pd$_{20}$Ge$_5$, anomalous temperature dependence in Raman spectra are reported in the ``$T_{2g}$'' mode of La \cite{HasegawaSan}. For this material, our results suggest that this $T_{2g}$ mode strongly interacts with conduction electrons, and the observed UD in $C_{44}$ can be naturally interpreted according to the discussion in \S \ref{harmoniccoupling}. For the optical phonon of the rare-earth dominated mode ($T_{1u}$) in filled-skutterudites, there are three degrees of freedom, namely $x,\ y$, and $z$. In order to discuss the anisotropy in the UD, it is desired to take into account these aspects of the system. For example, we must estimate the wave number dependence in the electron-phonon coupling $g$, which has a crucial role for the anisotropy in the UD. The results obtained in the present paper can be thought of as simplified but essential for describing the frequency dependence in the EC. Experimentally, there is an inconsistent result regarding the anisotropy of the UD in the different samples of PrOs$_4$Sb$_{12}$ \cite{Nakanishi}. This discrepancy should be investigated by further experimental efforts. 

At this stage, the question as to whether the rare-earth $T_{1u}$ mode is the relevant mode for the realization of the UD in filled skutterudites arises. It might be interesting to determine whether the temperature dependence of the $E_g$ modes in filled skutterudites is unusual, {\it e.g.,} whether softening occurs. It is important to carry out high-resolution Raman scattering experiments to clarify this point. Recently, Ogita {\it et al.}, observed the second-order Raman spectra in filled skutterudites (especially Sb compounds) as an anomalous property of the rare-earth mode \cite{Ogita2}. In the case in which the relevant mode is actually the rare-earth $T_{1u}$ mode in filled skutterudites, the mechanism of the anisotropy in the UD might be more complicated, as mentioned above. The $T_{1u}$ optical phonon should be investigated as a relevant optical phonon for the UD as a further theoretical study. Experiments in unfilled skutterudites with electric bands similar to, {\it e.g.}, LaOs$_4$Sb$_{12}$ are also desired in order to identify which mode is important for the UD. Applying this theory to optical phonons with even parity and comparing the energies of the optical phonons estimated by Raman scattering and the activation energy obtained by the ultrasound experiments for each materials will provide various interesting properties.  

In filled skutterudites, anomalous phonon contributions are observed in various quantities, especially in ROs$_4$Sb$_{12}$ (R=Pr, La, Sm, etc.). One reason why the effect is prominent in ROs$_4$Sb$_{12}$ may be that the size of the Sb$_{12}$ cage is the largest among existing filled skutterudites, including the rare-earth ion R. Another reason may be that the conduction electrons near the Fermi level in Os-compounds consist of a molecular orbital of the cage (A$_{1u}$) and the d-electron in Os. Unlike the Ru-compound, in which the levels of 4d electrons are deep compared to that of 5d or 3d electrons, so that their components near the Fermi level are negligible, the contributions from the d-electrons are much larger. It is thus expected that d-electrons play an important role in Os compounds. The latter point, however, requires more sophisticated analysis, because Iwasa {\it et al.,} recently observed the quasi-elastic peak even in Ru compounds \cite{iwasa2}. The observed width of the quasi-elastic peak is too large for UD to be realized in the MHz frequency range. In order to clarify the reason why the UD has been observed in Os-compounds but not in Ru-compounds, it may be necessary to carry out experiments in a wider frequency range. Note that if we assume the dominant electron-phonon coupling between the $A_{1u}$ molecular orbital and the $T_{1u}$ optical phonon is local and linear in the displacements, it is impossible for the $A_{1u}$ molecular orbital to interact with the $T_{1u}$ optical phonon due to symmetry.

Recently, nuclear magnetic resonance (NMR) experiments on KOs$_2$O$_6$ \cite{YoshidaKun} and LaOs$_4$Sb$_{12}$ \cite{NakaiNMR} were carried out. The results indicate that the relaxation is due to the quadrupolar coupling between the nuclei and the ionic motion (through a direct process) at K and La sites. In this case, $1/(T_1T)$ is given by a phonon spectral weight \cite{YoshidaKun}: $1/(T_1T)\propto -{\rm Im}D(z)/z$.
The results are interpreted by a strongly damped oscillator with an activation-type relaxation time. Although the energy scale of the ultrasonic and NMR experiments is of the same order as the magnitude ($\sim $ MHz), the results show no frequency dependence in $1/(T_1T)$. This might be due to the difference between the $q=0$ and $q$-integrated spectrum, or the difference of the phonon mode in the ultrasonic and NMR experiments, as mentioned above (the ultrasonic experiment for KOs$_2$O$_6$ has not yet been performed). Based on Fig. \ref{fig-phononLOW}, it is expected that $1/(T_1T)$ has frequency dependence. In the case in which the relaxation time is indeed an activation-type, {\it i.e.}, $\propto\exp(E_{\Delta}/T)$, the small change in $z$ does not indicate the same order change occurs in $T$, due to the exponential dependence. The precise origin of this problem is not yet understood and further experimental and theoretical studies are required.\cite{Dahm}

For the theoretical aspects, the approximated method used in the present paper becomes worse in the low-temperature regime, because $-\lim_{z\to0}{\rm Im}D(z)/z$ diverges as $T^{-2}$. This might cause the unphysical development of the soft mode. Intuitively, the divergence must be stopped at a certain temperature. However, it is important for this method to be able to capture the temperature dependence of the $-{\rm Im}D(z)$ for $z\ll \Omega_E$ at the intermediate temperature regions, where UD is observed.

 In our treatment of pseudo-fermion representations, it is straightforward to use higher-order interactions {\it e.g.,} $X^2$, $X^3$, $\cdots$ as interaction parts of the Hamiltonian. This is because the same argument can be applicable by simply replacing the matrix elements $X$ with $X^n$. The model including these higher-order electron-phonon couplings becomes important for ions with very large displacements, such as the rare-earth ion in filled skutterudites. The variations of hybridization between the f-electron and conduction electron by these ionic motions should be taken into account in a realistic model for the further clarification of electron-phonon systems. For the anharmonicity of the potential for the local phonon, it is straightforward to include higher-order terms such as $X^4$. Although the same line of discussion can be introduced in the present paper from an off-center potential, the presence of the off-center potential does not play a fundamental role with respect to our results. 

Finally, we mention the relationship between the present theory and our previous studies\cite{HattoriHirayamaMiyake1,HattoriHirayamaMiyake2}. In the present paper, we have discussed primarily the properties of a local phonon that interacts with conduction electrons. In our previous studies, however, we discussed the properties of electrons interacting with off-center configurations of an ion, which is an effective theory for the strong coupling limit of an electron-phonon system\cite{YotsuOgawa}. Further theoretical studies are needed in order to unify these aspects.

\section{Summary}\label{sec:sum}
In conclusion, we have investigated the acoustic and optical phonon spectra in a strongly coupled electron-phonon system at finite temperatures. We have applied the self-consistent ladder approximation, which was used successfully in the Coqblin-Schrieffer model with crystalline-electric-field states of Ce impurities, to the electron-phonon system, and have determined that the low-energy peak in the optical phonon spectral function develops at low-temperatures and causes the ultrasonic dispersion observed in filled-skutterudites and clathrate compounds. The temperature dependence of this peak shows activation type behaviors at the intermediate temperature regions, leading to the Debye-type formula for the ultrasonic dispersion with activation-type relaxation time. These results qualitatively explain the experimental results of the ultrasonic region observed in the filled skutterudites and related compounds.

\vspace{.3cm}
\section*{Acknowledgment}
\vspace{.3cm}
The authors would like to thank T. Goto, Y. Nemoto, Y. Nakanishi, and K. Kaneko for their fruitful discussions. The present study is supported in part by a Grant-in-Aid for Scientific Research in a Priority Area (No. 18027007) from MEXT, and by the 21st Century COE Program (G18) of the Japan Society for Promotion of Science (JSPS). One of the authors (K.H.) is also supported by a Research Fellowship for Young Scientists from JSPS.

\appendix
\section{Self-consistent Equations for the Self-Energies and Phonon Green Function in a Real Frequency}
In the Appendix, we list the self-consistent equation for the imaginary part of the retarded self-energies and the form of the retarded phonon Green function.

{\it Self-Energies}:\\
\begin{eqnarray}
{\rm Im}\bar{\Sigma}_A^{nm}(z)\!\!\!\!\!\!&=&\!\!\!\!2g^2\int \frac{dy}{\pi}n_F(y){\rm Im}G(-y)\nonumber\\\!\!\!\!\!\!\!&\times&\!\!\!\!\Big[\sum_{n'm'}X_{nm'}{\rm Im}\bar{\chi}_{GA}^{m'n'}(y+z)X_{n'm}\Big] \label{SigAreal}
\end{eqnarray}
and
\begin{eqnarray}
{\rm Im}\Sigma_G(z)\!\!\!\!&=&\!\!\!\!(1+e^{-z/T})g^2\int \frac{dy}{\pi}e^{-y/T}\sum_{nmn'm'}\nonumber\\
\!\!\!\!&\times&\!\!\!\!{\rm Im}\bar{A}_{mn'}(y+z){\rm Im}\bar{\chi}_{GA}^{m'n}(y)X_{nm}X_{n'm'}\nonumber\\
 \!\!\!\!&\times&\!\!\!\! \Big[ -\sum_l\int \frac{dx}{\pi}e^{-x/T}{\rm Im}\bar{A}_{ll}(x) \Big]^{-1},\label{SigFwaru}
\end{eqnarray}
with
\begin{eqnarray}
\!\!\!\!\!\!\!\!{\rm Im}{ \bar{\chi}_{GA}^0 }\ \!\!\!\!^{nm}(z)\!\!\!\!&=&\!\!\!\!\!\int \frac{dy}{\pi}n_F(y){\rm Im}G(y){\rm Im}\bar{A}_{nm}(z+y), \label{chi0real}
\end{eqnarray}
where $\bar{\chi}_{GA}$ in eq. (\ref{SigFwaru}) is obtained using eq. (\ref{laddMAT}). In eqs. (\ref{SigAreal}), (\ref{SigFwaru}), and (\ref{chi0real}), $n_F$ represents the Fermi distribution function. The function with an overbar is defined as $\bar{f}(x)\equiv f(x+\lambda)$, taking $\lambda\to \infty$ subject to the local constraint $\sum_n a^{\dag}_n a_n=1$. In eq. (\ref{SigFwaru}), we have divided the result, which is obtained by the analytic continuation of eq. (\ref{MatsuSigmaG}), by $\sum_n \langle a_n^{\dag}a_n\rangle$ \cite{Bickers}.
\vspace{0.1cm}\\

{\it Phonon Green Function}:
\begin{eqnarray}
{\rm Im}D(z)&=&-(1-e^{-z/T})\int \frac{dy}{\pi}e^{-y/T}\sum_{nn'mm'} \nonumber\\&\times&{\rm Im}\bar{A}_{mn'}(y+z){\rm Im}\bar{A}_{m'n}(y)X_{nm}X_{n'm'}\nonumber\\
 &\times& \Big[ -\sum_l\int \frac{dx}{\pi}e^{-x/T}{\rm Im}\bar{A}_{ll}(x) \Big]^{-1}.\label{ImD}
\end{eqnarray}
In eq. (\ref{ImD}), we have used the same prescription as that used for ${\rm Im}\Sigma_G$.

\end{document}